\documentstyle[11pt,epsfig]{article}
\begin{document}
\title{Extending the Broad Histogram Method for Continuous
  Systems\thanks{Oral presentation on CCP 1998, 2-5 September 1998,
    Granada, Spain}}
\author{JOSE D. MU\~NOZ$^{1,\dagger}$ and HANS J. HERRMANN$^{\ddagger}$\\
Institute for Computer Applications 1, University of Stuttgart,\\
Pfaffenwaldring 27, D-70569 Stuttgart, Germany\\
${}^{1}$Permanent address:
Dpto. de F\'{\i}sica, Univ. Nacional de Colombia,\\
Bogota D.C., Colombia\\
${}^{\dagger}${\footnotesize\rm E-mail: jdmunoz@ica1.uni-stuttgart.de}\\  
${}^{\ddagger}${\footnotesize\rm E-mail: hans@ica1.uni-stuttgart.de}\\
}
\maketitle
\begin{abstract}
We propose a way of extending the Broad Histogram Monte Carlo method
(BHMC) to systems with continuous degrees of freedom, and we apply
these ideas to investigate the three-dimensional XY-model.
Our method gives results in excellent agreement with Metropolis and
Histogram Monte Carlo simulations and calculates for the whole
temperature range $1.2<T<4.7$ using only $2$ times more computer
effort than the Histogram method for the range $2.1<T<2.2$. 
Our way of treatment is general, it can also be applied to other
systems with continuous degrees of freedom.
\end{abstract}

Usually one wants to calculate average values $<Q>_T$ of a certain
quantity $Q$ for a system in equilibrium at temperature $T$
according to, for instance, the canonical ensemble 
\begin{equation}
  \label{canonica1}
  <Q>_T={\sum_E <Q>_E g(E) \exp{(-E/k_BT)}  
         \over 
         \sum_Eg(E) \exp{(-E/k_BT)}},
\end{equation}
where $g(E)$ is the number of states with energy $E$, $<Q>_E$ denotes 
the micro-canonical average of $Q$ at energy $E$, and $k_B$ is the
Boltzmann constant (in the rest $k_B=1$ is taken). 
It is clear that given $g(E)$ and $<Q>_E$ one can calculate 
$<Q>_T$ at any desired temperature.

The Broad Histogram Monte Carlo Method (BHMC) developed by de Oliveira
{\it et. al.} \cite{BHMCstart,BHMCuC,BHMCisExact} calculates $g(E)$ and $<Q>_E$
directly. It chooses a microreversible protocol of allowed virtual movements in
the space of states of the system, and counts $N_{\mathrm up}(X)$
($N_{\mathrm dn}(X)$) as the number of allowed movements that
increases (decreases) the energy of the configuration $X$ by a fixed
amount $\Delta E_{\mathrm fix}$.
These movements are virtual, in the sense that they are never
performed. They are introduced only to estimate $g(E)$ and should not
be mixed with the dynamics employed to take the samples. 
It has been shown \cite{BHMCisExact} that
the total number of possible ways to go up, summed over all states
with energy $E$, equals the total number of ways to go down, summed
over all states with energy $E+\Delta E_{\mathrm fix}$, i.e.\ 
\begin{equation}
 \label{basic}
   g(E) <N_{\mathrm up}(E)> = g(E+\Delta E_{\mathrm fix})
   <N_{\mathrm dn}(E+\Delta E_{\mathrm fix})> \quad ,
\end{equation}
where $<N_{\mathrm up}(E)>$ ($<N_{\mathrm dn}(E)>$) is the
micro-canonical average of $N_{\mathrm up}(X)$ ($N_{\mathrm dn}(X)$)
at energy $E$. 
By taking logarithms of both sides of Eq.\
(\ref{basic}) and by subsequent dividing by $\Delta E_{\mathrm fix}$
one can obtain expressions for $\ln g(E+\Delta E_{\mathrm fix})- \ln
g(E)$ and $\beta(E) \equiv {d \ln g(E) / d E}$ that can be either
added up or integrated numerically to obtain $\ln g(E)$.
The micro-canonical averages involved in the method can be calculated
either analytically or by Monte Carlo simulations. In this case four
histograms are required, i.e.\ $N_{up}(E)$, $N_{dn}(E)$, $Q(E)$ and
the number of visits $V(E)$. 
This procedure gives accurate results with small computer
efforts for many discrete systems \cite{BHMCstart,BHMCisExact}.

To extend the BHMC method to continuous systems we propose to
choose a protocol of virtual random movements in the space
of states of the system, such that for each allowed movement the
probability to perform it equals the probability to revert it, i.e.\
    \begin{eqnarray}
      \label{microequilibrado}
      P{\small (X_{\mathrm old} \to X_{\mathrm new})} =
      P{\small (X_{\mathrm new} \to X_{\mathrm old})} \quad .
    \end{eqnarray}
Next, let us define the probability density function (p.d.f.) $f^X (\Delta E)$
i.e.\ the probability to obtain an energy change between $\Delta 
E$ and $\Delta E+d\Delta E$ starting from the configuration $X$. We propose
\begin{eqnarray}
  \label{NupNdn}
  N_{\mathrm up}(X) \equiv f^X (\Delta E_{fix}) & ; &
  N_{\mathrm dn}(X) \equiv f^X (- \Delta E_{fix}) \quad .
\end{eqnarray}
Eq.\  (\ref{basic}) is still valid,
and the method proceeds as before \cite{BHMCcont}.

To test our ideas we chose the XY-model \cite{XY_Adler} with the Hamiltonian
\begin{equation}
  \label{Hamiltonian}
  \mathcal{H}= -{\it J} \sum_{<{\it i j}>} \vec{\sigma}_{\it i} \cdot
  \vec{\sigma}_{\it j} 
  = -{\it J} \sum_{ < {\it  ij} > } \cos (\theta_{\it i} - \theta_{\it
    j}) \quad ,
\end{equation}
where the summations $<ij>$ are taken over all pairs of
nearest-neighbor sites, $\theta_{\it i}$ denotes the planar angle of
the spin at site $i$ and $J$ is the maximal energy per bond (in the
rest $J=1$). As protocol of virtual random movements, we take at
random one site
$i$ with angle ${\theta_i}_{\mathrm old}$ and choose a new angle
${\theta_i}_{\mathrm new}$ for it with uniform probability on $[-\pi ,+\pi]$. 
We obtain \cite{BHMCcont}
\begin{eqnarray}
  \label{fDeltaEi}
 f^X{\scriptstyle (\Delta E)}= {1 \over N} \sum_i f^i
 {\scriptstyle (\Delta E)}  &;&   
 f^i {\scriptstyle (\Delta E)}= \left\{ \begin{array} {cc} 
 {1 \over \pi} {1 \over \sqrt{A_i^2-(\varepsilon_i+\Delta E)^2}} 
 & {\scriptstyle |\Delta E+\varepsilon_i|<A_i}
 \\ 0 & {\scriptstyle otherwise}
\end{array} \right. \quad ,
\end{eqnarray}
where $N$ denotes the number of sites $i$, and 
\begin{eqnarray}
   \varepsilon_i \equiv - \sum_{j} \cos {({\theta_i}_{\mathrm old} - \theta_j)} &;&
   A_i {\scriptstyle \equiv \sqrt{\left( \sum_{j} \cos \theta_j
   \right)^2 + \left( \sum_{j} \sin \theta_j \right)^2}}
\end{eqnarray}
The summation over $j$ is always performed on all nearest neighbors of
site $i$. Finally $N_{\mathrm up}(X)$ ($N_{\mathrm dn}(X)$) equals the
average of $f^i (\Delta E_{\mathrm fix})$ ($f^i (- \Delta E_{\mathrm
  fix})$) on all sites. 

To estimate the micro-canonical averages  we binned the
whole energy range of positive temperatures  
(i.e.\  energies per bond between $-1.0$ and $0.0$) in windows of size
$\Delta E_{fix}$ and performed a micro-canonical simulation on each
window as follows \cite{BHMCuC}. Starting from a configuration inside
the window one spin was rotated to a new angle chosen at random with
uniform probability over the interval $[-\pi,\pi]$. 
If the new system energy falls inside the window, the
change is performed, otherwise it is rejected. It is clear that this
sampling process maintains a detailed balance condition with equal
probabilities for the old an the new configurations and, therefore, it
samples all the states inside the window with the same probability. An
entire lattice sweep was performed by repeating this procedure on all
sites. 

\begin{figure}[hbt!]
  \includegraphics[width=0.48\textwidth, bb= 45 45 535 445]{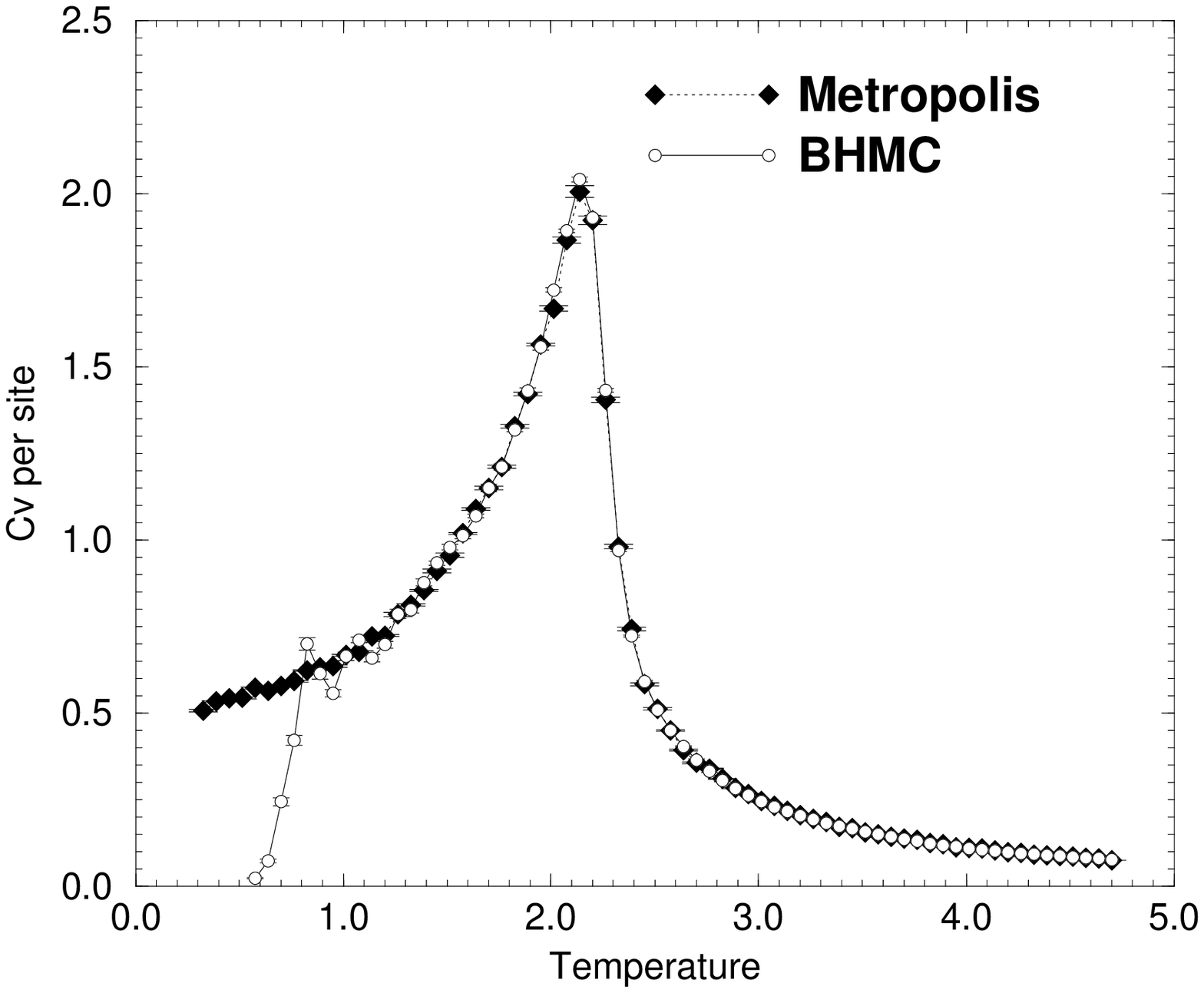}
  \hfill
  \includegraphics[width=0.48\textwidth, bb= 45 45 535 445]{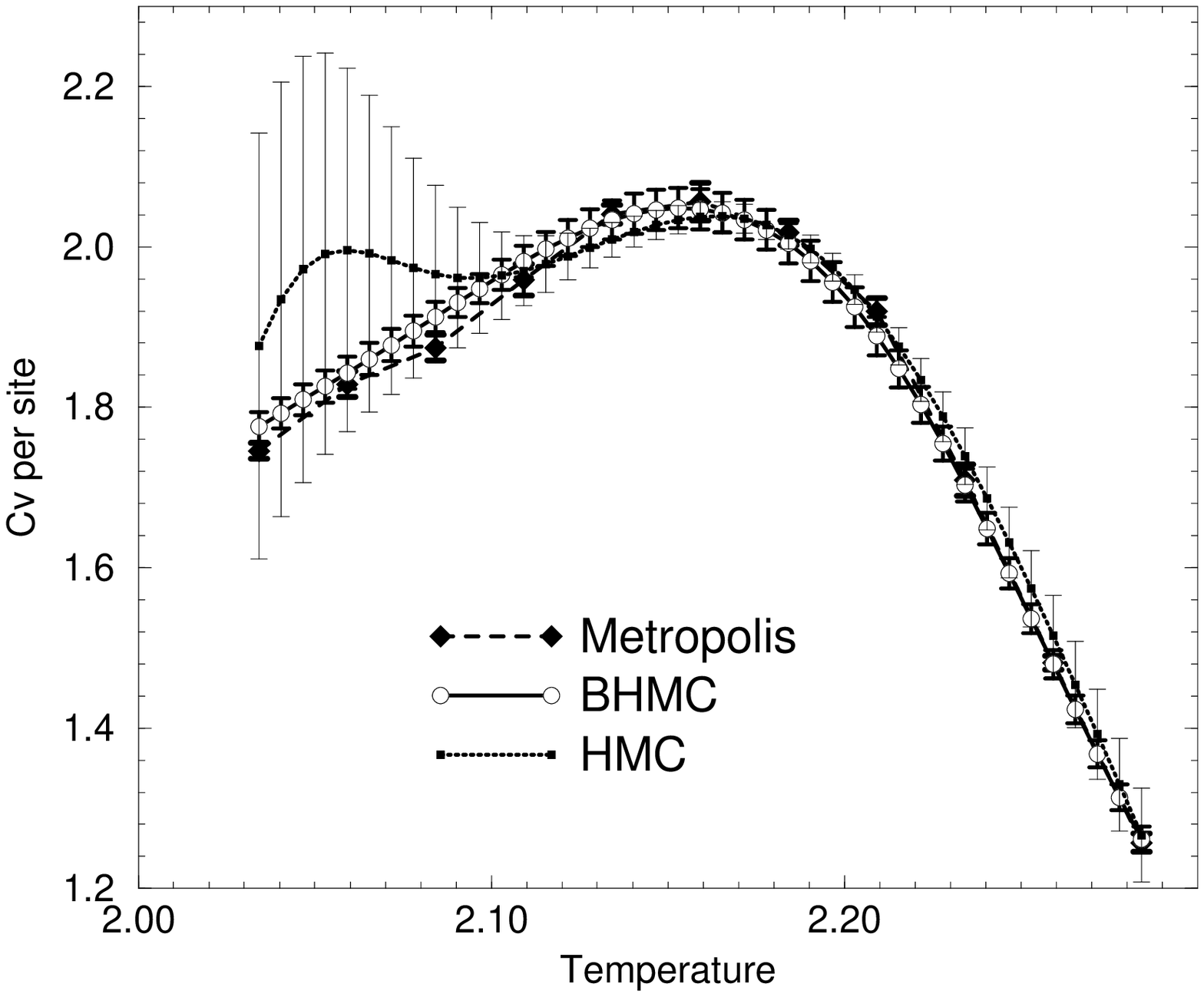}
  \caption{\footnotesize Specific heat for the  SC ${10 \times 10 \times 10}$
    XY-model obtained from the BHMC method (circles), the HMC method
    (small squares) and Metropolis simulations (filled diamonds). The
    error bars on the left (right) side correspond to $1.0$ ($3.5$,
    i.e.\ 99\% confidence level) standard deviations for eight runs.}
  \label{Cv}
\end{figure}

Fig.\ \ref{Cv} compares the specific heat we obtained
for the SC ${10 \times 10 \times 10}$ XY-model with periodic boundary
conditions by using the BHMC method (circles), the Histogram Monte
Carlo method (HMC) \cite{Salzburg,FerrSwend} (squares)
and Metropolis \cite{Metropolis} simulations (diamonds).
For the BHMC method $\Delta E_{fix}=6.0$ ($\Delta
E_{fix}=0.002$ in energy-per-bond units) was chosen.

The HMC method took $188$ sec.\ per run, a Metropolis for one
temperature point $28.5$ sec.\  and the BHMC method $383$ sec.\ per
run on a Digital Alpha workstation, i.e.\ only $2.0$ times more
computer effort than the HMC method and less than $14$ Metropolis
simulations.
It can be observed that all methods coincide on the temperature range
$2.1<T<2.2$, i.e.\ the classical range of validity for this HMC
simulation \cite{HMCerror},
but the BHMC method is still precise on the whole temperature
range $1.2<T<4.7$. 

In conclusion, our proposal to apply the BHMC method to continuous
systems gives excellent agreement with Metropolis and HMC
simulations. The strategy proposed is completely general, it can also
be applied to other systems with continuous degrees of freedom.

\end{document}